\documentclass[12pt]{article}
\usepackage{amsmath,amssymb}
\usepackage{amscd}
\newcommand{\eqdef}{\stackrel{\text{def}}{=}}
\newcommand{\n}{\nonumber\\}
\newcommand{\bm}{\boldsymbol}

\allowdisplaybreaks[4]

\begin{document}

\baselineskip=20pt

\newfont{\elevenmib}{cmmib10 scaled\magstep1}
\newcommand{\preprint}{
 \vspace*{-20mm}
    \begin{flushleft}
     \elevenmib Yukawa\, Institute\, Kyoto\\
   \end{flushleft}\vspace{-1.3cm}
   \begin{flushright}\normalsize \sf
     YITP-10-17\\
     April 2010
   \end{flushright}}
\newcommand{\Title}[1]{{\baselineskip=26pt
   \begin{center} \Large \bf #1 \\ \ \\ \end{center}}}
\newcommand{\Author}{\begin{center}
   \large \bf  Ryu Sasaki${}^a$, Satoshi Tsujimoto${}^b$
   and Alexei Zhedanov${}^{b,c}$
    \end{center}}
\newcommand{\Address}{\begin{center}
    ${}^a$ Yukawa Institute for Theoretical Physics,\\
     Kyoto University, Kyoto 606-8502, Japan\\
     ${}^b$ Department of Applied Mathematics and Physics,\\
Graduate School of Informatics,
Kyoto University, Kyoto 606-8501, Japan\\
${}^c$ Donetsk Institute for Physics and Technology,\\
Donetsk 83114, Ukraine
   \end{center}}
\newcommand{\Accepted}[1]{\begin{center}
   {\large \sf #1}\\ \vspace{1mm}{\small \sf Accepted for Publication}
   \end{center}}

\preprint
\thispagestyle{empty}
\bigskip\bigskip\bigskip

\Title{Exceptional Laguerre and Jacobi polynomials and the corresponding 
potentials through Darboux-Crum transformations}
\Author

\Address

\begin{abstract}
Simple derivation is presented of the four families of
infinitely many shape invariant Hamiltonians corresponding to 
the exceptional Laguerre and Jacobi polynomials.
Darboux-Crum transformations are applied
to connect the well-known shape invariant Hamiltonians of the
radial oscillator and the Darboux-P\"oschl-Teller potential to the 
shape invariant potentials of Odake-Sasaki.
Dutta and Roy derived the two lowest members
of the exceptional Laguerre polynomials by this method.
The method is expanded to its full generality and many other ramifications,
including the aspects of generalised Bochner problem and the bispectral
property of the exceptional orthogonal polynomials,  are discussed.
\end{abstract}

%
%
%

\section{Introduction}
\label{sec:intro}
\setcounter{equation}{0}

Here we construct in an elementary way the 
four sets of infinitely many shape invariant Hamiltonians and the
corresponding exceptional ($X_{\ell}$) polynomials \cite{os16,os19}. The idea is quite simple.
We start with a prepotential $W_\ell(x;\bm{\lambda})$ and define  a pair
of factorised Hamiltonians $\mathcal{H}^{(\pm)}_\ell(\bm{\lambda})$ 
which are intertwined by the the
Darboux-Crum transformations \cite{darboux,crum} in terms of 
$A_\ell(\bm{\lambda})$ and $A_\ell(\bm{\lambda})^\dagger$:
\begin{align}
\mathcal{H}^{(+)}_\ell(\bm{\lambda})&
\eqdef A^\dagger_\ell(\bm{\lambda})A_\ell(\bm{\lambda}),\quad 
\mathcal{H}^{(-)}_\ell(\bm{\lambda})\eqdef A_\ell(\bm{\lambda}) A^\dagger_\ell(\bm{\lambda}), 
\quad
\ell=1,2,\ldots,\\
 A_\ell(\bm{\lambda}) \mathcal{H}^{(+)}_\ell(\bm{\lambda})&
 = \mathcal{H}^{(-)}_\ell(\bm{\lambda}) A_\ell(\bm{\lambda}),\quad
A^\dagger_\ell(\bm{\lambda}) \mathcal{H}^{(-)}_\ell(\bm{\lambda})
= \mathcal{H}^{(+)}_\ell(\bm{\lambda}) A^\dagger_\ell(\bm{\lambda}),
\label{intertwine}\\
A_\ell(\bm{\lambda}) &\eqdef \frac{d}{dx}-\frac{dW_\ell(x;\bm{\lambda})}{dx},
\quad A^\dagger_\ell(\bm{\lambda})=-\frac{d}{dx}-\frac{dW_\ell(x;\bm{\lambda})}{dx}.
\end{align}
Here $\bm{\lambda}$ stands for the set of parameters of the theory.
The prepotential  $W_\ell(x;\bm{\lambda})$ is so chosen that
$\mathcal{H}^{(+)}_\ell(\bm{\lambda})$ is the well-known shape invariant \cite{genden} 
Hamiltonian
of  the radial oscillator \cite{infhul,susyqm} potential or the Darboux-P\"oschl-Teller (DPT) 
potential \cite{dpt} and $\mathcal{H}^{(-)}_\ell(\bm{\lambda})$ is the Hamiltonian of the recently
derived shape invariant potentials of Odake-Sasaki \cite{os16,os19,os18,hos}.
The essential property  of the prepotential to achieve the above goal 
is that $e^{\pm W_\ell(x;\bm{\lambda})}$
is {\em not square integrable\/}. 
We know that $\mathcal{H}^{(+)}_\ell(\bm{\lambda})$ has a well defined groundstate.
Thus these two functions $e^{\pm W_\ell(x;\bm{\lambda})}$ cannot correspond
to the groundstates  of  $\mathcal{H}^{(+)}_\ell(\bm{\lambda})$ 
or $\mathcal{H}^{(-)}_\ell(\bm{\lambda})$.
That is, the groundstates of $\mathcal{H}^{(\pm)}_\ell(\bm{\lambda})$ are not annihilated by
$A_\ell(\bm{\lambda})$ or $A^\dagger_\ell(\bm{\lambda})$. 
This means that $\mathcal{H}^{(+)}_\ell(\bm{\lambda})$ 
and $\mathcal{H}^{(-)}_\ell(\bm{\lambda})$ are 
{\em exactly iso-spectral\/} including the groundstates:
\begin{align}
\left.
\begin{array}{l}
 \mathcal{H}^{(+)}_\ell(\bm{\lambda})\phi_{\ell, n}^{(+)}(x;\bm{\lambda}) =
 \mathcal{E}_{\ell,n}^{(+)}(\bm{\lambda})\phi_{\ell, n}^{(+)}(x;\bm{\lambda}) \\[6pt]
 \mathcal{H}^{(-)}_\ell(\bm{\lambda})\phi_{\ell, n}^{(-)}(x;\bm{\lambda})=
 \mathcal{E}_{\ell,n}^{(-)}(\bm{\lambda})\phi_{\ell, n}^{(-)}(x;\bm{\lambda})
\end{array}
\right\},
\quad  \mathcal{E}_{\ell,n}^{(+)}(\bm{\lambda})= \mathcal{E}_{\ell,n}^{(-)}(\bm{\lambda})>0,\\
\ell=1,2,\ldots,\quad n=0,1,2,\ldots, .
\end{align}
By construction $\mathcal{H}^{(+)}_\ell(\bm{\lambda})$ is shape invariant and exactly solvable.
That is, the set of eigenvalues $\{\mathcal{E}_{\ell,n}^{(+)}(\bm{\lambda})\}$ 
and the corresponding eigenfunctions
$\{\phi_{\ell, n}^{(+)}(x;\bm{\lambda})\}$ are exactly known.
Throughout this paper we choose all the eigenfunctions to be real.
They form a {\em complete set of orthogonal functions\/}:
\begin{equation}
\int \phi_{\ell, m}^{(+)}(x;\bm{\lambda}) \phi_{\ell, n}^{(+)}(x;\bm{\lambda})dx 
=h_{\ell,n}(\bm{\lambda}) \delta_{m\, n},\quad h_{\ell,n}(\bm{\lambda})>0.
\label{genortho}
\end{equation}
Thanks to the intertwining relations \eqref{intertwine} the eigenfunctions 
$\{\phi_{\ell, n}^{(-)}(x;\bm{\lambda})\}$ of the partner
Hamiltonian $\mathcal{H}^{(-)}_\ell(\bm{\lambda})$ are obtained from the well-known
eigenfunctions $\{\phi_{\ell, n}^{(+)}(x;\bm{\lambda})\}$ of the radial oscillator or the DPT
Hamiltonian $\mathcal{H}^{(\pm)}_\ell(\bm{\lambda})$ by the Darboux-Crum 
transformation in terms of
$ A_\ell(\bm{\lambda})$
\begin{equation}
\phi_{\ell, n}^{(-)}(x;\bm{\lambda})= A_\ell(\bm{\lambda}) \phi_{\ell, n}^{(+)}(x;\bm{\lambda})
,\quad
\ell=1,2,\ldots,\quad n=0,1,2,\ldots, 
\label{minplurel}
\end{equation}
and vice versa:
\begin{equation}
\phi_{\ell, n}^{(+)}(x;\bm{\lambda})
=\frac{A^\dagger_\ell(\bm{\lambda})}{\mathcal{E}_{\ell,n}^{(+)}(\bm{\lambda})} \phi_{\ell, n}^{(-)}(x;\bm{\lambda}),
\quad
\ell=1,2,\ldots,\quad n=0,1,2,\ldots, .
\end{equation}
Of course $\{\phi_{\ell, n}^{(-)}(x;\bm{\lambda})\}$ is another {\em complete\/} set of   {\em orthogonal functions\/}:
\begin{align}
&\int \phi_{\ell, m}^{(-)}(x;\bm{\lambda}) \phi_{\ell, n}^{(-)}(x;\bm{\lambda})dx
= \int A_\ell(\bm{\lambda})\phi_{\ell, m}^{(+)}(x;\bm{\lambda})\cdot
A_\ell(\bm{\lambda}) \phi_{\ell, n}^{(+)}(x;\bm{\lambda})dx\n
&\qquad=\int \phi_{\ell, m}^{(+)}(x;\bm{\lambda})\cdot
A_\ell(\bm{\lambda})^\dagger A_\ell(\bm{\lambda}) \phi_{\ell, n}^{(+)}(x;\bm{\lambda})
\n
&\qquad={\mathcal{E}_{\ell,n}^{(+)}(\bm{\lambda})} 
\int \phi_{\ell, m}^{(+)}(x;\bm{\lambda}) \phi_{\ell, n}^{(+)}(x;\bm{\lambda})dx
={\mathcal{E}_{\ell,n}^{(+)}(\bm{\lambda})}h_{\ell,n}(\bm{\lambda})\delta_{m\,n}.
\label{genorthomin}
\end{align}
These imply the {\em completeness\/} of the exceptional orthogonal polynomials
and the above relationship \eqref{minplurel} provides the formula relating the exceptional 
orthogonal polynomials to the classical orthogonal polynomials ({\em i.e,} 
the Laguerre or Jacobi
polynomials) as shown in (2.1) and (2.3) of \cite{hos}.
The orthogonality \eqref{genorthomin} corresponds to the integration formulas derived in \S7 of \cite{hos}.  
These will be demonstrated in detail in subsequent sections.

The above requirements lead to the following general form of the prepotential
$W_\ell(x;\bm{\lambda})$:
\begin{equation}
W_\ell(x;\bm{\lambda})=\widetilde{w}_0(x;\bm{\lambda}+\ell\bm{\delta})
+\log\xi_\ell(\eta(x);\bm{\lambda}),
\end{equation}
in which $\widetilde{w}_0(x;\bm{\lambda})$
 is obtained by changing the sign of one term of
the prepotential $w_0(x;\bm{\lambda})$ corresponding to the radial oscillator or the DPT potential.
Here $\bm{\delta}$ is the shift of the parameters.
The change of the sign ensures  the non-square integrability of $e^{\pm W_\ell(x;\bm{\lambda})}$.
The additional term is the logarithm of the degree $\ell$ eigenpolynomial, the Laguerre or  Jacobi polynomial with
twisted parameters or arguments, 
introduced by Odake-Sasaki  \cite{os16, os19,hos}. 

Before going to the details in the subsequent sections, let us make a few remarks about the background.
This type of approach of deriving a new exactly solvable Hamiltonian from a known one in
terms of  Darboux-Crum \cite{darboux, crum} transformations has a long history 
and various aspects \cite{nieto, bagsam, junkroy}. 
The method we are concerned in this paper is, in its essence, 
based on an alternative factorisation of an exactly solvable Hamiltonian (plus a constant),
{\em e.g.}, the radial oscillator and the DPT potential.
Some refer to those newly found Hamiltonians as ``conditionally exactly solvable". 
Junker and Roy \cite{junkroy} discussed an example of an alternative factorisation
of the radial oscillator Hamiltonian by using the confluent hypergeometric
function ${}_1F_1$, which could encompass the results of the L1 exceptional orthogonal 
polynomials \cite{os16} if the parameters and settings are properly chosen.
After the introduction of the $X_1$ Laguerre polynomials by Gomez-Ullate et
 al \cite{gomez1} and
Quesne \cite{quesne,quesne2} and the $X_\ell$ ($\ell=1,2,\ldots$) polynomials by Odake-Sasaki
\cite{os16,os19}, Roy and his collaborator \cite{duttaroy}
derived the $X_1$ and $X_2$ Laguerre polynomials of the L1 type in this way. 
This will be mentioned in a later section.
A recent report by Gomez-Ullate eta al \cite{gomez2} has small overlap with the present work.
So far as we are aware of, a ``conditionally exactly solvable" treatment of the fully general 
(non-symmetric) DPT potential does not exist. 
Therefore, the present derivation of the J1 and J2 exceptional Jacobi polynomials
from the classical Jacobi polynomials based on the Darboux-Crum transformation is new.

The exceptional orthogonal polynomials were originally introduced in \cite{gomez1} by extending Bochner's theorem \cite{bochner} for Sturm-Liouville problems.
The characterisation of the exceptional orthogonal polynomials as polynomial solutions
of Sturm-Liouville type equations under generalised Bochner problems will be discussed in section \ref{sec:bochner}. We will show that the exceptional Laguerre and Jacobi polynomials
have the {\em bispectral property\/}  \cite{GH2}. 
On top of the Sturm-Liouville type equation, the $X_\ell$
polynomials satisfy $4\ell+1$ recursion relations \eqref{rec_exc_Jac} and \eqref{hat_L_rec},
which could be rewritten as an eigenvalue equation \eqref{KEVP} of a semi-infinite 
matrix $K$ which acts on the label $n$ of the exceptional polynomials 
$\hat{P}_n(x)$ and $\hat{L}_n(x)$. Simple interpretation of the bispectral property is provided as 
the characteristic feature of the invariant polynomial subspaces of the Sturm-Liouville type
operator.

This paper is organised as follows. In section \ref{exceplag}, the two types (L1 and L2)
of infinitely many  exceptional Laguerre polynomials are derived by simple Darboux-Crum transformations connecting them with the Laguerre polynomials. 
The two types (J1 and J2)
of infinitely many  exceptional Jacobi polynomials are derived in a similar way by using
Darboux-Crum transformations connecting them with the Jacobi  polynomials. 
The aspects of the generalised Bochner problems, in particular, the bispectral property will be discussed in section \ref{sec:bochner}. Based on the so-called Darboux transformations of the 
orthogonal polynomials, that is the Christoffel and Geronimus transformations
of the Jacobi and Laguerre polynomials \cite{Bateman, Zhe1}, the bispectral property of the exceptional 
polynomials are derived elementarily. 
In other words, the $X_\ell$ polynomials are shown to satisfy  $4\ell+1$ term recurrence relations, 
which are the generalisation of the well known three term recurrence relations.
The final section is for comments and discussions.
It provides a simple proof of shape invariance of the Hamiltonians of the exceptional
orthogonal polynomials. It is shown that they also possess the creation/annihilation operators inherited from the radial oscillator/DPT Hamiltonian systems. 
It is stressed that the Hamiltonians of the radial oscillator/DPT potentials admit infinitely
many non-singular factorisations, which induce the Darboux-Crum transformations.
The extended `three term recurrence' relations for the exceptional orthogonal polynomials are also discussed from the Darboux-Crum transformations point of view.

\section{Exceptional Laguerre polynomials}
\label{exceplag}
\setcounter{equation}{0}
Here we will derive the L1 and L2 exceptional Laguerre
 polynomials as well as the
corresponding Hamiltonians, that is the potentials.

\subsection{Radial Oscillator}
\label{radial}
Let us start with the radial oscillator with $\bm{\lambda}=g>0$ and $\bm{\delta}=1$:
\begin{align}
  &w_0(x;g)\eqdef -\tfrac12 x^2+g\log x, \qquad 0<x<\infty,\\
  &\mathcal{H}_R^{(+)}(g)=p^2+x^2+\frac{g(g-1)}{x^2}-1-2g.
\end{align}
It is trivial to verify the shape invariance \cite{genden}:
\begin{equation}
\mathcal{H}_R^{(-)}(g)=\mathcal{H}_R^{(+)}(g+1)+4,\quad \mathcal{E}_1(g)=4.
\end{equation}
Its eigenvalues and eigenfunctions are
\begin{align}
  &\mathcal{E}_n(g)=4n,\\
  &\phi_n(x;g)=P_n(\eta;g)\,e^{w_0(x;g)},\quad \eta\equiv \eta(x)\eqdef x^2,\quad
  \,P_n(x;g)=L_n^{(g-\frac12)}(x),
  \end{align}
  in which $L_n^{(\alpha)}(x)$ is the Laguerre polynomial satisfying the differential equation
\begin{equation}
  x\partial_x^2L_n^{(\alpha)}(x)
  +(\alpha+1-x)\partial_xL_n^{(\alpha)}(x)
  +nL_n^{(\alpha)}(x)=0.
  \label{Ldiffeq}
\end{equation}
It should be stressed that the groundstate wavefunction $\phi_0(x;g)=e^{w_0(x;g)}=e^{-x^2/2}x^g$ 
is square integrable and it provides the orthogonality measure
of the Laguerre polynomials;
\begin{equation}
\int_0^\infty e^{2w_0(x;g)}P_m(x^2;g)P_n(x^2;g)dx=h_n(g) \delta_{m\,n},
\quad
h_n(g)\eqdef\frac{1}{2\,n!}\,\Gamma(n+g+\tfrac12).
  \label{normL0}
\end{equation}
The radial oscillator Hamiltonian system is exactly solvable in the Heisenberg picture, too
\cite{os7}. The exact annihilation/creation operators are obtained as the positive/negative energy parts
of the Heisenberg operator solution (see, for example, (3.8) of \cite{os7}):
\begin{equation}
  a^{(\pm)}
  =\Bigl(\bigl(\frac{d}{dx}\mp x\bigr)^2-\frac{g(g-1)}{x^2}\Bigr)\!\bigm/4.
  \label{acradosci}
\end{equation}
The action of these operators are 
\begin{equation}
  a^{(-)}\phi_n(x;g)=-(n+g-\frac12)\phi_{n-1}(x;g),\quad
  a^{(+)}\phi_n(x;g)=-(n+1)\phi_{n+1}(x;g).
  \label{actionacradosci}
\end{equation}

\subsection{L1 and L2 exceptional Laguerre polynomials}
\label{L1L2lag}
Here we derive the L1 and L2 exceptional Laguerre polynomials.
For each positive integer $\ell=1,2,\ldots$, let us consider the pair of Hamiltonians  
$\mathcal{H}^{(+)}_\ell(g)$ and $\mathcal{H}^{(-)}_\ell(g)$ corresponding to the following 
prepotentials ($\eta\equiv\eta(x)\eqdef x^2$)
\begin{align}
\text{L1:}\quad
W_\ell(x;g)&\eqdef\frac{x^2}{2}+(g+\ell-1)\log x +\log\xi_\ell(\eta(x);g),\quad g>1/2,
\label{L1prepot}\\
&\hspace{50mm}\xi_\ell(\eta;g)\eqdef L_\ell^{(g+\ell-\frac32)}(-\eta),
\label{L1xi}\\
\text{L2:}\quad W_\ell(x;g)&\eqdef-\frac{x^2}{2}-(g+\ell)\log x +\log\xi_\ell(\eta(x);g),\quad g>-1/2,
\label{L2prepot}\\
&\hspace{50mm}\xi_\ell(\eta;g)\eqdef L_\ell^{(-g-\ell-\frac12)}(\eta),
\label{L2xi}
\\
\mathcal{H}^{(+)}_\ell(g)&\eqdef A^\dagger_\ell(g)A_\ell(g),\qquad \qquad \quad
\mathcal{H}^{(-)}_\ell(g)\eqdef A_\ell(g) A^\dagger_\ell(g),\\
A_\ell(g) &\eqdef \frac{d}{dx}-\frac{dW_\ell(x;g)}{dx},\qquad 
A^\dagger_\ell(g)=-\frac{d}{dx}-\frac{dW_\ell(x;g)}{dx}.
\end{align}
By simple calculation using the differential equation for the Laguerre polynomial
\eqref{Ldiffeq}, the Hamiltonian $\mathcal{H}^{(+)}_\ell(g)$ is shown to be equal to the radial oscillator
with $g\to g+\ell-1$ for L1 and with $g\to g+\ell+1$ for L2 up to an additive constant:
\begin{align}
\text{L1:}\quad \mathcal{H}^{(+)}_\ell(g)
&=p^2+x^2+\frac{(g+\ell-1)(g+\ell-2)}{x^2}+2g+6\ell-1,
\label{radosci2}\\
&=\mathcal{H}^{(+)}_R(g+\ell-1)+2(2g+4\ell-1),\\
\text{L2:}\quad \mathcal{H}^{(+)}_\ell(g)
&=p^2+x^2+\frac{(g+\ell)(g+\ell+1)}{x^2}+2(g-\ell)-1,
\label{radosci3}\\
&=\mathcal{H}^{(+)}_R(g+\ell+1)+2(2g+1).
\end{align}
The partner Hamiltonians are
\begin{align}
\text{L1:}\quad \mathcal{H}^{(-)}_\ell(g)
&=p^2+x^2+\frac{(g+\ell)(g+\ell-1)}{x^2}+2(g-\ell)-3\n
&\qquad +2\left(\frac{\partial_x\xi_\ell(\eta;g)}{\xi_\ell(\eta;g)}\right)
\left(2\frac{(g+\ell-1)}{x}+\frac{\partial_x\xi_\ell(\eta;g)}{\xi_\ell(\eta;g)}\right),
\label{partnerradosci2}\\
\text{L2:}\quad \mathcal{H}^{(-)}_\ell(g)
&=p^2+x^2+\frac{(g+\ell)(g+\ell-1)}{x^2}+2(g+3\ell)+1\n
&\qquad +2\left(\frac{\partial_x\xi_\ell(\eta;g)}{\xi_\ell(\eta;g)}\right)
\left(-2\left(x+\frac{(g+\ell)}{x}\right)+\frac{\partial_x\xi_\ell(\eta;g)}{\xi_\ell(\eta;g)}\right).
\label{partnerradosci3}
\end{align} 
Up to  additive constants, the above Hamiltonians \eqref{partnerradosci2} 
and \eqref{partnerradosci3} are equal to the
Hamiltonians of the L1 and L2 exceptional orthogonal polynomials derived by 
Odake-Sasaki \cite{os16,hos}:
\begin{align}
\mathcal{H}^{\text{O-S}}_\ell(g)&\eqdef p^2+\left(\frac{dw_\ell(x;g)}{dx}\right)^2+\frac{d^2w_\ell(x;g)}{dx^2},\\
&w_\ell(x;g)\eqdef -\frac{x^2}{2}+(g+\ell)\log x +\log\frac{\xi_\ell(\eta;g+1)}{\xi_\ell(\eta;g)},\\
\text{L1:}\quad \mathcal{H}^{(-)}_\ell(g)&=\mathcal{H}^{\text{O-S}}_\ell(g)+2(2g+4\ell-1),\\
\text{L2:}\quad \mathcal{H}^{(-)}_\ell(g)&=\mathcal{H}^{\text{O-S}}_\ell(g)+2(2g+1).
\end{align}
The definition of $\xi_\ell(\eta;g)$ for the L1 and L2 Odake-Sasaki cases are the same as those given in 
\eqref{L1xi} and \eqref{L2xi}.

Let us note that  $e^{W_\ell(x;g)}$ is {\em not} square integrable at $x=\infty$ and
$e^{-W_\ell(x;g)}$ is {\em not} square integrable at $x=0$ for the L1 case, 
whereas  for the L2 case $e^{W_\ell(x;g)}$ is {\em not} square integrable at $x=0$ and
$e^{-W_\ell(x;g)}$ is {\em not} square integrable at $x=\infty$.
In both cases the prepotential
$W_\ell(x;g)$ \eqref{L1prepot} and \eqref{L2prepot}  are regular in the interval $0<x<\infty$.
For this, it is enough to show that $\xi_\ell(\eta(x);g)$ does not have a zero in
$0<x<\infty$. In fact we have, 
\begin{align}
&\text{L1:}\quad  \xi_{\ell}(\eta(x);g)
  =\sum_{k=0}^{\ell}\frac{(g+\ell+k-\frac12)_{\ell-k}}{k!\,(\ell-k)!}
  \,x^{2k}>0,\\
&\text{L2:}\quad    (-1)^\ell \xi_{\ell}(\eta(x);g)
  =\sum_{k=0}^{\ell}\frac{(g+\frac12)_{\ell-k}}{k!\,(\ell-k)!}
  \,x^{2k}>0,
\end{align}
as shown in (2.39) of \cite{os18}. 
Thus we find that $e^{\pm W_\ell(x;g)}$ cannot be the groundstates of the 
Hamiltonians $\mathcal{H}^{(\pm)}_\ell(g)$. 
However, we know quite well that $\mathcal{H}^{(+)}_\ell(g)$, being the radial oscillator Hamiltonian, 
has a well-defined groundstate. This means that the partner Hamiltonian $\mathcal{H}^{(-)}_\ell(g)$, 
thus the Hamiltonian of the L1  and L2 exceptional Laguerre polynomials, 
are exactly iso-spectral to the radial oscillator Hamiltonian
$\mathcal{H}^{(+)}_\ell(g)$, which have the following eigenvalues and
the corresponding eigenfunctions:
\begin{alignat}{2}
&\text{L1:}\quad \mathcal{E}_{\ell,n}^{(+)}(g)=4n+2(2g+4\ell-1), \quad
&\phi_{\ell,n}^{(+)}(x;g)=e^{-\frac{x^2}{2}}x^{g+\ell-1}L_n^{(g+\ell-\frac32)}(\eta),
\label{L1plus}\\
&\text{L2:}\quad \mathcal{E}_{\ell,g}^{(+)}(g)=4n+2(2g+1),\quad
&\phi_{\ell,n}^{(+)}(x;g)=e^{-\frac{x^2}{2}}x^{g+\ell+1}L_n^{(g+\ell+\frac12)}(\eta).
\label{L2plus}
\end{alignat}
The intertwining relation \eqref{intertwine} implies the simple expressions of the eigenfunctions of 
the partner Hamiltonians $\mathcal{H}_\ell^{(-)}$ in terms of $A_\ell(g)$:
\begin{align}
&\text{L1:}\qquad
\mathcal{E}_{\ell,n}^{(-)}(g)=4n+2(2g+4\ell-1), \qquad\qquad\qquad \mathcal{E}_{\ell,n}^{\text{O-S}}(g)=4n,\\
&\qquad\quad\ \phi_{\ell,n}^{\text{O-S}}(x;g)=\phi_{\ell,n}^{(-)}(x;g)= 
\left(\frac{d}{dx}-\frac{dW_\ell(x;g)}{dx}\right)e^{-\frac{x^2}{2}}x^{g+\ell-1}L_n^{(g+\ell-\frac32)}(\eta)\n
&\qquad\quad\ \phantom{\phi_{\ell,n}^{\text{O-S}}(x;g)}= 2\frac{e^{-\frac{x^2}{2}}x^{g+\ell}}{ \xi_{\ell}(\eta;g)}
\left(- \xi_{\ell}(\eta;g+1)L_n^{(g+\ell-\frac32)}(\eta)+ \xi_{\ell}(\eta;g)\partial_\eta L_n^{(g+\ell-\frac32)}(\eta)\right),
\label{L1eig}\\
&\text{L2:}\qquad
\mathcal{E}_{\ell,n}^{(-)}(g)=4n+2(2g+1),\qquad\qquad\qquad\quad  \mathcal{E}_{\ell,n}^{\text{O-S}}(g)=4n,\\
&\qquad\quad\ \phi_{\ell,n}^{\text{O-S}}(x;g)=\phi_{\ell,n}^{(-)}(x;g)= 
\left(\frac{d}{dx}-\frac{dW_\ell(x;g)}{dx}\right)e^{-\frac{x^2}{2}}x^{g+\ell+1}L_n^{(g+\ell+\frac12)}(\eta)\n
&\qquad\quad\ \phantom{\phi_{\ell,n}^{\text{O-S}}(x;g)}= 2\frac{e^{-\frac{x^2}{2}}x^{g+\ell}}{ \xi_{\ell}(\eta;g)}
\left((g+\frac12)\xi_{\ell}(\eta;g+1)L_n^{(g+\ell+\frac12)}(\eta)\right.\n
&\hspace*{95mm}\left. 
+ \eta\,\xi_{\ell}(\eta;g)\partial_\eta L_n^{(g+\ell+\frac12)}(\eta)\right).
\label{L2eig}
\end{align}
These are to be compared with the explicit expressions of the exceptional Laguerre
polynomials  $P_{\ell,n}(\eta;g)$ derived in \cite{hos}:
\begin{align} 
\phi_{\ell,n}(x;g)&=\psi_{\ell}(x;g)P_{\ell,n}(\eta;g),\quad 
\psi_{\ell}(x;g)\eqdef\frac{e^{-\frac{x^2}{2}}x^{g+\ell}}{\xi_{\ell}(\eta;g)},\quad 
 \xi_{\ell}(\eta;g)\eqdef
  \left\{
  \begin{array}{ll}
  L_{\ell}^{(g+\ell-\frac32)}(-\eta)&:\text{L1}\\
  L_{\ell}^{(-g-\ell-\frac12)}(\eta)&:\text{L2}
  \end{array}\right.,\n[4pt]
 & P_{\ell,n}(\eta;g)\eqdef
  \left\{
  \begin{array}{ll}
  \xi_{\ell}(\eta;g+1)L_n^{(g+\ell-\frac32)}(\eta)
  -\xi_{\ell}(\eta;g)\partial_{\eta}
  L_n^{(g+\ell-\frac32)}(\eta)&:\text{L1}\\[2pt]
  (n+g+\frac12)^{-1}\bigl((g+\frac12)
  \xi_{\ell}(\eta;g+1)L_n^{(g+\ell+\frac12)}(\eta)\\
  \phantom{(n+g+\frac12)^{-1}\bigl(}\ \quad
  +\eta\xi_{\ell}(\eta;g)\partial_{\eta}L_n^{(g+\ell+\frac12)}(\eta)
  \bigr)&:\text{L2},
  \end{array}\right.. \hspace{8mm} (\text{O-S}2.1)\nonumber
\end{align}
The final expressions of the eigenfunctions \eqref{L1eig} and \eqref{L2eig} are the same as (2.1) of \cite{hos} up to a multiplicative constant. 
Use is made of the identities of the Laguerre polynomials (E.11) and (E.12)  of \cite{hos}.
Thus we have derived the Hamiltonians as well as the eigenfunctions, that is, 
 the L1 and L2 exceptional Laguerre polynomials and the weight functions, from those of 
 the radial oscillator by the Darboux-Crum transformations.

\section{Exceptional Jacobi polynomials}
\label{excepjac}
\setcounter{equation}{0}

Here we will derive the J1 and J2 exceptional Jacobi polynomials as well as the
corresponding Hamiltonians, that is the potentials.

\subsection{Trigonometric DPT potential}
\label{trigDPT}
The trigonometric DPT \cite{dpt} potential has two parameters $\bm{\lambda}=(g,h)$, $g>0$, $h>0$ and $\bm{\delta}=(1,1)$,
\begin{align}
w_0(x;g,h)\eqdef g\log\sin x+h\log\cos x,\quad 0<x<\frac{\pi}{2},\\
\mathcal{H}_{DPT}^{(+)}(g,h)=p^2+\frac{g(g-1)}{\sin^2x}+\frac{h(h-1)}{\cos^2x}-(g+h)^2.
\label{tridpt}
\end{align}
It is trivial to verify the shape invariance \cite{genden}:
\begin{equation}
\mathcal{H}_{DPT}^{(-)}(g,h)=\mathcal{H}_{DPT}^{(+)}(g+1,h+1)+4(g+h+1),
\quad \mathcal{E}_1(g,h)=4(g+h+1).
\end{equation}
Its eigenvalues and eigenfunctions are:
\begin{align}
  &\mathcal{E}_n(g,h)=4n(n+g+h),\\
  &\phi_n(x;g,h)
  =P_n(\eta;g,h)\,e^{w_0(x;g,h)},\quad \eta\equiv \eta(x)\eqdef\cos2x,
\quad
  P_n(x;g,h)=P_n^{(g-\frac12,h-\frac12)}(x),
  \label{jacpoly}
\end{align}
in which $P_n^{(\alpha,\beta)}(x)$ is the Jacobi polynomial satisfying the
second order differential equation
\begin{equation}
  (1-x^2)\partial_x^2P_n^{(\alpha,\beta)}(x)
  +\bigl(\beta-\alpha-(\alpha+\beta+2)x\bigr)\partial_x P_n^{(\alpha,\beta)}(x)
  +n(n+\alpha+\beta+1)P_n^{(\alpha,\beta)}(x)=0.
  \label{Jdiffeq}
\end{equation}
The groundstate wavefunction $\phi_0(x;g,h)=e^{w_0(x;g,h)}=(\sin x)^g(\cos x)^h$ 
is square integrable and it provides the orthogonality measure
of the Jacobi polynomials:
\begin{align}
\int_0^{\pi/2} e^{2w_0(x;g,h)}P_m(\eta(x);g,h)P_n(\eta(x);g,h)dx=h_n(g,h) \delta_{m\,n},\\
 h_n(g,h)\eqdef \frac{\Gamma(n+g+\frac12)\Gamma(n+h+\frac12)}
  {2\,n!(2n+g+h)\Gamma(n+g+h)}.
\end{align}
The trigonometric DPT potential is also exactly solvable in the Heisenberg picture \cite{os7}.
The annihilation and creation operators are (see, for example, (3.28) of \cite{os7}): 
\begin{equation}
  a^{\prime(\pm)}/2=a^{(\pm)}2\sqrt{\mathcal{H}'}
  =\pm \sin 2x\frac{d}{dx}+\cos 2x\,\sqrt{\mathcal{H}'}
  +\frac{\alpha^2-\beta^2}{\sqrt{\mathcal{H}'}\pm 1},
\end{equation}
in which
\begin{equation}
\mathcal{H}'\eqdef \mathcal{H}_{DPT}+(g+h)^2,\quad 
\alpha\eqdef g-\frac12,\quad \beta\eqdef h-\frac12.
\end{equation}
When applied to the eigenvector $\phi_n$ \eqref{jacpoly} as 
$\mathcal{E}_n(g,h)+(g+h)^2=(2n+g+h)^2$, we obtain (see, for example, (3.29)
and (3.30) of \cite{os7}):
\begin{align}
  a^{\prime(-)}\phi_n(x;g,h)/2
  &=-\sin 2x\frac{d\phi_n(x;g,h)}{dx}+(2n+g+h)\cos 2x\,\phi_n(x;g,h)\n
  &\qquad +\frac{\alpha^2-\beta^2}{2n+\alpha+\beta}\,\phi_n(x;g,h)\n
  &=\frac{4(n+\alpha)(n+\beta)}{2n+\alpha+\beta}\,\phi_{n-1}(x;g,h),
  \label{PTdown}\\
  a^{\prime(+)}\phi_n(x;g,h)/2
  &=\phantom{-}\sin 2x\frac{d\phi_n(x;g,h)}{dx}+(2n+g+h)\cos 2x\,\phi_n(x;g,h)\n
 &\qquad +\frac{\alpha^2-\beta^2}{2n+\alpha+\beta+2}\,\phi_n(x;g,h)\n
  &=\frac{4(n+1)(n+\alpha+\beta+1)}{2n+\alpha+\beta+2}\,\phi_{n+1}(x;g,h).
  \label{PTup}
\end{align}

\subsection{J1 and J2 exceptional Jacobi polynomials}
\label{J1jac}
Here we derive the J1 and J2 exceptional Jacobi polynomials.
As explained in \cite{os18,hos}, the exceptional J1 and J2 orthogonal polynomials are 
`mirror images' of each other, reflecting the parity property 
$P_n^{(\alpha,\beta)}(-x)=(-1)^nP_n^{(\beta,\alpha)}(x)$ of 
the Jacobi polynomial. Here we present both cases in parallel so that the structure 
of these polynomials can be better understood by comparison.

For each positive integer $\ell=1,2,\ldots$, let us consider the pair of Hamiltonians  
$\mathcal{H}^{(+)}_\ell(g,h)$ and $\mathcal{H}^{(-)}_\ell(g,h)$ corresponding to the 
following prepotential ($\eta\equiv\eta(x)\eqdef \cos2x$)
\begin{align}
\text{J1:}\quad
W_\ell(x;g,h)&\eqdef (g+\ell-1)\log \sin x -(h+\ell)\log\cos x+ \log\xi_\ell(\eta;g,h),
\label{J1prepot}\\
&\hspace{20mm}\xi_\ell(\eta;g,h)\eqdef P_\ell^{(g+\ell-\frac32,-h-\ell-\frac12)}(\eta),\quad g>h>0,
\label{J1xi}\\
\text{J2:}\quad
W_\ell(x;g,h)&\eqdef -(g+\ell)\log \sin x +(h+\ell-1)\log\cos x+ \log\xi_\ell(\eta;g,h),
\label{J2prepot}\\
&\hspace{20mm}\xi_\ell(\eta;g,h)\eqdef P_\ell^{(-g-\ell-\frac12, h+\ell-\frac32)}(\eta),\quad h>g>0,
\label{J2xi}\\
\mathcal{H}^{(+)}_\ell(g,h)&\eqdef A^\dagger_\ell(g,h)A_\ell(g,h),\qquad \qquad \quad
\mathcal{H}^{(-)}_\ell(g,h)\eqdef A_\ell(g,h) A^\dagger_\ell(g,h),\\
A_\ell(g,h) &\eqdef \frac{d}{dx}-\frac{dW_\ell(x;g,h)}{dx},\qquad 
A^\dagger_\ell(g,h)=-\frac{d}{dx}-\frac{dW_\ell(x;g,h)}{dx}.
\end{align}
By simple calculation using the differential equation for the Jacobi polynomial
\eqref{Jdiffeq}, we obtain the trigonometric DPT potential for $\mathcal{H}^{(+)}_\ell(g,h)$
up to an additive constant:
\begin{align}
\text{J1:}\quad
\mathcal{H}^{(+)}_\ell(g,h)
&=p^2+\frac{(g+\ell-1)(g+\ell-2)}{\sin^2x}+\frac{(h+\ell)(h+\ell+1)}{\cos^2x}-(2\ell+g-h-1)^2,
\label{trigdpt2}\\
&=\mathcal{H}_{DPT}^{(+)}(g+\ell-1,h+\ell+1)+(2g+4\ell-1)(2h+1),\\
\text{J2:}\quad
\mathcal{H}^{(+)}_\ell(g,h)
&=p^2+\frac{(g+\ell)(g+\ell+1)}{\sin^2x}+\frac{(h+\ell-1)(h+\ell-2)}{\cos^2x}-(2\ell+h-g-1)^2,
\label{trigdpt3}\\
&=\mathcal{H}_{DPT}^{(+)}(g+\ell+1,h+\ell-1)+(2h+4\ell-1)(2g+1)
\end{align}
The partner Hamiltonians are
\begin{align}
\text{J1:}\quad
\mathcal{H}^{(-)}_\ell(g,h)
&=p^2+\!\frac{(g+\ell)(g+\ell-1)}{\sin^2x}+\frac{(h+\ell)(h+\ell-1)}{\cos^2x}\n
&\ +2\left(\frac{\partial_x\xi_\ell(\eta;g,h)}{\xi_\ell(\eta;g,h)}\right)
\left(2\left((g+\ell-1)\cot x+(h+\ell)\tan x\right)+\frac{\partial_x\xi_\ell(\eta;g,h)}{\xi_\ell(\eta;g,h)}\right)\n
&-(2\ell+g-h-1)^2+8\ell(\ell+g-h-1),
\label{partnertrigdptJ1}\\
\text{J2:}\quad
\mathcal{H}^{(-)}_\ell(g,h)
&=p^2+\!\frac{(g+\ell)(g+\ell-1)}{\sin^2x}+\frac{(h+\ell)(h+\ell-1)}{\cos^2x}\n&\ 
+2\left(\frac{\partial_x\xi_\ell(\eta;g,h)}{\xi_\ell(\eta;g,h)}\right)
\left(-2\left((g+\ell)\cot x+(h+\ell-1)\tan x\right)+\frac{\partial_x\xi_\ell(\eta;g,h)}{\xi_\ell(\eta;g,h)}\right)\n
&-(2\ell+h-g-1)^2+8\ell(\ell+h-g-1).
\label{partnertrigdptJ2}
\end{align} 
Up to  additive constants, the above Hamiltonian \eqref{partnertrigdptJ1} 
and \eqref{partnertrigdptJ2} are equal to the
Hamiltonian of the J1  and J2 exceptional Jacobi polynomials derived by 
Odake-Sasaki \cite{os16,hos}:
\begin{align}
\mathcal{H}^{\text{O-S}}_\ell(g,h)&\eqdef p^2
+\left(\frac{dw_\ell(x;g,h)}{dx}\right)^2+\frac{d^2w_\ell(x;g,h)}{dx^2},\\
w_\ell(x;g,h)&\eqdef (g+\ell)\log\sin x +(h+\ell)\log\cos x 
+\log\frac{\xi_\ell(\eta;g+1,h+1)}{\xi_\ell(\eta;g,h)},\\
\text{J1:}\quad
\mathcal{H}^{(-)}_\ell(g,h)&=\mathcal{H}^{\text{O-S}}_\ell(g,h)+(2g+4\ell-1)(2h+1),\\
\text{J2:}\quad
\mathcal{H}^{(-)}_\ell(g,h)&=\mathcal{H}^{\text{O-S}}_\ell(g,h)+(2h+4\ell-1)(2g+1).
\end{align}
The definition of $\xi_\ell(\eta;g)$ for the J1 and J2 Odake-Sasaki cases are the same as those given in 
\eqref{J1xi} and \eqref{J2xi}.
Again let us note that  $e^{W_\ell(x;g,h)}$ is {\em not} square integrable at $x=\pi/2$ and
$e^{-W_\ell(x;g,h)}$ is {\em not} square integrable at $x=0$ for the J1 case, 
whereas for the J2 case $e^{W_\ell(x;g,h)}$ is {\em not} square integrable at $x=0$ and
$e^{-W_\ell(x;g,h)}$ is {\em not} square integrable at $x=\pi/2$. But the prepotentials
$W_\ell(x;g,h)$ \eqref{J1prepot} and \eqref{J2prepot}  are regular in the interval $0<x<\pi/2$.
For this, it is enough to show that $\xi_\ell(\eta(x);g,h)$ does not have a zero in
$0<x<\pi/2$. In fact we have, 
\begin{align}
&\text{J1:}\quad
  (-1)^{\ell}\xi_{\ell}(\eta(x);g,h)= \frac{(h+\frac12)_{\ell}}{\ell!}\sum_{k=0}^{\ell}
  \frac{(\ell-k+1)_k(g-h+\ell-1)_k}{k!\,(h+\ell-k+\frac12)_k}
  (\cos x)^{2k}>0,\\
&\text{J2:}\quad
    (-1)^{\ell}\xi_{\ell}(\eta(x);g,h)=\frac{(g+\frac12)_{\ell}}{\ell!}\sum_{k=0}^{\ell}
  \frac{(\ell-k+1)_k(h-g+\ell-1)_k}{k!\,(g+\ell-k+\frac12)_k}
  (\sin x)^{2k}>0,
\end{align}
as shown in (2.40) of \cite{os18}. 
Thus we find that $e^{\pm W_\ell(x;g,h)}$ cannot be the groundstates of the 
Hamiltonians $\mathcal{H}^{(\pm)}_\ell(g,h)$. However, we know well that 
$\mathcal{H}^{(+)}_\ell(g,h)$, being the trigonometric DPT Hamiltonian, has
a well-defined groundstate. This means that the partner Hamiltonian $\mathcal{H}^{(-)}_\ell(g,h)$, 
thus the Hamiltonian of the J1 and J2
exceptional Jacobi polynomials, are exactly iso-spectral to the trigonometric DPT Hamiltonians
$\mathcal{H}^{(+)}_\ell(g,h)$, which have the following eigenvalues and
the corresponding eigenfunctions:
\begin{align}
\text{J1:}\quad
\mathcal{E}_{\ell,n}^{(+)}(g,h)&=4n(n+g+h+2\ell)+(2g+4\ell-1)(2h+1),\\
\phi_{\ell,n}^{(+)}(x;g,h)&=(\sin x)^{g+\ell-1}(\cos x)^{h+\ell+1}
P_n^{(g+\ell-\frac32,h+\ell+\frac12)}(\eta),\\
  \text{J2:}\quad
  \mathcal{E}_{\ell,n}^{(+)}(g,h)&=4n(n+g+h+2\ell)+(2h+4\ell-1)(2g+1),\\
\phi_{\ell,n}^{(+)}(x;g,h)&=(\sin x)^{g+\ell+1}(\cos x)^{h+\ell-1}P_n^{(g+\ell+\frac12,h+\ell-\frac32)}(\eta).
\end{align}
The intertwining relation \eqref{intertwine} implies the simple expressions of the eigenfunctions 
of the partner Hamiltonians $\mathcal{H}_\ell^{(-)}$ in terms of $A_\ell(g)$:
\begin{align}
\text{J1:}\quad
&\mathcal{E}_{\ell,n}^{(-)}(g,h)=4n(n+g+h+2\ell)+(2g+4\ell-1)(2h+1),\\
&\qquad\qquad \mathcal{E}_{\ell,n}^{\text{O-S}}(g,h)=4n(n+g+h+2\ell),\\
&\phi_{\ell,n}^{\text{O-S}}(x;g,h)=\phi_{\ell,n}^{(-)}(x;g,h)
= \left(\frac{d}{dx}-\frac{dW_\ell(x;g,h)}{dx}\right)\n
&\hspace*{60mm} \times (\sin x)^{g+\ell-1}(\cos x)^{h+\ell+1}
P_n^{(g+\ell-\frac32,h+\ell+\frac12)}(\eta)\n
&\phantom{\phi_{\ell,n}^{\text{O-S}}(x;g,h)}= 
-2\frac{(\sin x)^{g+\ell}(\cos x)^{h+\ell}}{ \xi_{\ell}(\eta;g,h)}
\left((h+\frac12) \xi_{\ell}(\eta;g+1,h+1)P_n^{(g+\ell-\frac32, h+\ell+\frac12)}(\eta)\right.\n
&\hspace*{55mm}\left.+ (1+\eta)\xi_{\ell}(\eta;g,h)
\partial_\eta P_n^{(g+\ell-\frac32, h+\ell+\frac12)}(\eta)\right),
\label{J1sol}\\
\text{J2:}\quad
&\mathcal{E}_{\ell,n}^{(-)}(g,h)=4n(n+g+h+2\ell)+(2h+4\ell-1)(1+2g),\\
&\qquad\qquad \mathcal{E}_{\ell,n}^{\text{O-S}}(g,h)=4n(n+g+h+2\ell),\\
&\phi_{\ell,n}^{\text{O-S}}(x;g,h)=\phi_{\ell,n}^{(-)}(x;g,h)
= \left(\frac{d}{dx}-\frac{dW_\ell(x;g,h)}{dx}\right)\n
&\hspace*{60mm} \times (\sin x)^{g+\ell+1}(\cos x)^{h+\ell-1}P_n^{(g+\ell+\frac12,h+\ell-\frac32)}(\eta)\n
&\phantom{\phi_{\ell,n}^{\text{O-S}}(x;g,h)}= -2\frac{(\sin x)^{g+\ell}(\cos x)^{h+\ell}}{ \xi_{\ell}(\eta;g,h)}
\left((g+\frac12) \xi_{\ell}(\eta;g+1,h+1)P_n^{(g+\ell+\frac12, h+\ell-\frac32)}(\eta)\right.\n
&\hspace*{55mm}\left.- (1-\eta)\xi_{\ell}(\eta;g,h)
\partial_\eta P_n^{(g+\ell+\frac12, h+\ell-\frac32)}(\eta)\right).
\label{J2sol}
\end{align}
These are to be compared with the explicit expressions of the exceptional Jacobi
polynomials  $P_{\ell,n}(\eta;g,h)$ derived in \cite{hos}:
\begin{align} 
&\phi_{\ell,n}(x;g,h)=\psi_{\ell}(x;g,h)P_{\ell,n}(\eta;g,h),\quad 
\psi_{\ell}(x;g,h)\eqdef\frac{(\sin x)^{g+\ell}(\cos x)^{h+\ell}}{ \xi_{\ell}(\eta;g,h)},\n
&\hspace{40mm}   \xi_{\ell}(\eta;g,h)\eqdef
  \left\{
  \begin{array}{ll}
  P_{\ell}^{(g+\ell-\frac32,-h-\ell-\frac12)}(\eta),\ \ g>h>0&:\text{J1}\\
  P_{\ell}^{(-g-\ell-\frac12,h+\ell-\frac32)}(\eta),\ \ h>g>0&:\text{J2}
  \end{array}\right. ,\n
 &   P_{\ell,n}(\eta;g,h)\eqdef
  \left\{
  \begin{array}{ll}
  (n+h+\frac12)^{-1}\bigl(
  (h+\frac12)\xi_{\ell}(\eta;g+1,h+1)
  P_n^{(g+\ell-\frac32,h+\ell+\frac12)}(\eta)&\\
  \phantom{(n+h+\frac12)^{-1}\bigl(}
  +(1+\eta)\xi_{\ell}(\eta;g,h)
  \partial_{\eta}P_{n}^{(g+\ell-\frac32,h+\ell+\frac12)}(\eta)\bigr)
  &:\text{J1}\\[2pt]
  (n+g+\frac12)^{-1}\bigl(
  (g+\frac12)\xi_{\ell}(\eta;g+1,h+1)
  P_n^{(g+\ell+\frac12,h+\ell-\frac32)}(\eta)&\\
  \phantom{(n+g+\frac12)^{-1}\bigl(}
  -(1-\eta)\xi_{\ell}(\eta;g,h)
  \partial_{\eta}P_{n}^{(g+\ell+\frac12,h+\ell-\frac32)}(\eta)\bigr)
  &:\text{J2}
  \end{array}\right.. \hspace{2mm} (\text{O-S}2.3)\nonumber
\end{align}
The final expressions \eqref{J1sol} and \eqref{J2sol} are the same as (2.3) of \cite{hos} up to a multiplicative constant. 
Use is made of the identity of the Jacobi polynomials (E.22) of \cite{hos}.
Thus we have derived the Hamiltonians as well as the eigenfunctions
of the J1 and J2 exceptional Jacobi polynomials from those of the trigonometric 
DPT by the Darboux-Crum transformations.

The exceptional Laguerre and Jacobi polynomials satisfy a second order
linear differential equation in the {\em entire complex $\eta$ plane\/}:
\begin{equation}
  \widetilde{\mathcal{H}}^{\text{O-S}}_{\ell}(\bm{\lambda})
  P_{\ell,n}(\eta;\bm{\lambda})
  =\mathcal{E}_{\ell,n}(\bm{\lambda})P_{\ell,n}(\eta;\bm{\lambda}),\quad
  \mathcal{E}_{\ell,n}(\bm{\lambda})
  =\mathcal{E}_n(\bm{\lambda}+\ell\bm{\delta}).
  \label{htildeeq}
\end{equation}
For later use we give the explicit form of the second order Fuchsian differential operator 
$\widetilde{\mathcal H}^{\text{O-S}}_\ell$ 
which was given in (3.5) of \cite{hos}:
\begin{align}
  \widetilde{\mathcal{H}}_{\ell}^{\text{O-S}}(\bm{\lambda})
  &=-4\Bigl(c_2(\eta)\frac{d^2}{d\eta^2}
  +\bigl(c_1(\eta,\bm{\lambda}+\ell\bm{\delta})-2c_2(\eta)
  \partial_{\eta}\log\xi_{\ell}(\eta;\bm{\lambda})\bigr)\frac{d}{d\eta}\n
  &\phantom{=-4\Bigl(}\quad
  +2d_1(\bm{\lambda})\frac{c_2(\eta)}{d_2(\eta)}
  \frac{\partial_{\eta}\xi_{\ell}(\eta;\bm{\lambda}+\bm{\delta})}
  {\xi_{\ell}(\eta;\bm{\lambda})}
  +\frac14\,\widetilde{\mathcal{E}}_{\ell}(\bm{\lambda}+\bm{\delta})\Bigr),
  \label{newttildeeq}
\end{align}
in which $c_1$, $c_2$, $d_1$, $d_2$ and $\widetilde{\mathcal E}_\ell$ are given by
\begin{align}
  c_1(\eta,\bm{\lambda})&\eqdef\left\{
  \begin{array}{ll}
  g+\tfrac12-\eta&:\text{L}\\
  h-g-(g+h+1)\eta&:\text{J}
  \end{array}\right.,
  \quad
  c_2(\eta)\eqdef\left\{
  \begin{array}{ll}
  \eta&:\text{L}\\
  1-\eta^2&:\text{J}
  \end{array}\right.,
  \label{c1,c2}\\[2pt]
   d_1(\bm{\lambda})&\eqdef\left\{
  \begin{array}{cl}
  1&:\text{L1}\\
  g+\frac12&:\text{L2,J2}\\
  h+\frac12&:\text{J1}
  \end{array}\right.,
  \qquad\qquad\qquad\
  d_2(\eta)\eqdef\left\{
  \begin{array}{ll}
  1&:\text{L1}\\
  -\eta&:\text{L2}\\
  \mp(1\pm \eta)&:\text{J1/J2}
  \end{array}\right.,
  \label{d12def}\\[2pt]
  \widetilde{\mathcal{E}}_{\ell}(\bm{\lambda})&\eqdef\left\{
  \begin{array}{ll}
  \mp 4\ell&:\text{L1/L2}\\
  4\ell(\ell\pm g\mp h-1)&:\text{J1/J2}
  \end{array}\right..
  \label{En,Etl}
\end{align}
\section{Generalised Bochner problem: bispectral property}
\label{sec:bochner}
\setcounter{equation}{0}
The exceptional Laguerre polynomials L1 and L2 as well as the exceptional Jacobi polynomials J1 
and J2 belong to complete orthogonal families of functions 
(the completeness follows from the well known properties of the Darboux process; 
indeed, the Darboux transformation which does not generate new eigenstates preserves 
the completeness of transformed system of eigenfunctions as solutions of the 
self-adjoint Schr\"odinger equation).
These polynomials, however, do not belong to the ordinary families of orthogonal polynomials because polynomials of the first $\ell-1$ degrees are absent in these systems \cite{gomez1}. 
Hence these polynomials do not satisfy 3-term recurrence relation which is a characteristic property of nondegenerate orthogonal polynomials (see, e.g. \cite{Ismail}).
Nevertheless, as we will show, the exceptional polynomials J1, J2, L1 and L2 do satisfy $4\ell+1$-term recurrence relations, 
i.e. they can be considered as eigenvectors of a semi-infinite matrix $K$ having $4\ell+1$ diagonals. 
In this sense the considered exceptional polynomials possess 
a very important bispectral property \cite{GH2}: 
{\em they  are simultaneously eigenfunctions of a Sturm-Liouville operator and a matrix $K$}. 

\subsection{Bispectrality of the exceptional Jacobi polynomials }

We consider exceptional orthogonal polynomials of the J1-type 
(the type J2 can be considered in the same manner). 
For simplicity of presentation we slightly change the previous notation, denoting the exceptional J1 polynomials as 
$\hat P_n(x)$ and introducing standard parameters  
$a\eqdef g+\ell-\frac32,\:b\eqdef h+\ell+\frac12$. 
We also use $x$ for the sinusoidal coordinate $\eta$ and denote 
$\pi(x)= \xi_{\ell}(x;g,h)=P_{\ell}^{(a,-b)}(x)$ 
which is a polynomial of degree $\ell$. This polynomial will play a crucial role in the following. 
Then formula (O-S2.3) for the J1 polynomials can be presented in the form
\begin{equation}
\hat P_n(x) =
\pi(x) \left((1+x) P_n'^{(a,b)}(x) + b P_n^{(a,b)}(x) \right) -
\pi'(x) (1+x) P_n^{(a,b)}(x),
\label{exc_Jac} 
\end{equation} 
which is equal to $P_{\ell,n}(x;g,h)$ in (O-S2.3) up to a multiplicative constant.
Recall that the  Jacobi polynomials $P_n^{(a,b)}(x)$ satisfy the three-term recurrence relation
\begin{equation}
A_nP_{n+1}^{(a,b)}(x) + B_n P_n^{(a,b)}(x) + C_n P_{n-1}^{(a,b)}(x) = x P_n^{(a,b)}(x), 
\label{rec_J}
\end{equation}
with
\begin{align} 
 A_n&= \frac{2(n+1)(n+a+b+1)}{(2n+a+b+1)(2n+a+b+2)}, \quad
 B_n = \frac{b^2-a^2}{(2n+a+b)(2n+a+b+2)}, \n
C_n &= \frac{2(n+a)(n+b)}{(2n+a+b)(2n+a+b+1)} .
\nonumber
\end{align}
They are orthogonal on the interval $[-1,1]$
\begin{equation}
\int_{-1}^1 P_n^{(a,b)}(x) P_m^{(a,b)}(x) (1-x)^{a}(1+x)^{b} dx = h_n  \: \delta_{nm}, 
\label{J_ort} 
\end{equation}
where
\begin{equation*}
{h}_n\eqdef \frac{2^{a+b+1}\Gamma(n+a+1)\Gamma(n+b+1)}
  {n!(2n+a+b+1)\Gamma(n+a+b+1)}
\end{equation*}
is the normalization constant.
The exceptional J1 polynomials are orthogonal on the same interval
\begin{equation}
(\hat P_n,\hat P_m)\eqdef \int_{-1}^1 \hat P_n(x) \hat P_m(x) \hat w(x) dx = \hat h_n  \: \delta_{nm}, 
\label{hat_J_ort} 
\end{equation}
with the weight function
\begin{equation}
\hat w(x) =  \frac{(1-x)^{a+1}(1+x)^{b-1}}{\pi^2(x)} 
\label{w_Jac_b} 
\end{equation}
and some nonzero normalization coefficients $\hat h_n$ 
(in fact, these coefficients can easily be connected with the coefficients $h_n$, 
however we need not their explicit expressions here).

Using elementary properties of the Jacobi polynomials \cite{Bateman}, or (2.24) of \cite{hos}, we also have 
\begin{equation}
\hat P_n(x) = (b+n) \pi(x)
P_n^{(a+1,b-1)}(x) - \pi'(x) (1+x) P_n^{(a,b)}(x).
\label{exc_Jac_11} 
\end{equation}
Moreover, there are classical formulas 
\cite{Bateman} 
\begin{equation}
(2n+a+b+1)(1+x)P_n^{(a,b)}(x) =2(n+1) P_{n+1}^{(a,b-1)}(x)+2(n+b) P_n^{(a,b-1)}(x),
 \label{CT_Jac}
\end{equation}
and 
\begin{equation}
(2n+a+b+1)P_n^{(a,b)}(x) =(n+a+b+1)P_{n}^{(a+1,b)}(x)-(n+b)P_{n-1}^{(a+1,b)}(x).
\label{GT_Jac} 
\end{equation}
These formulas have a simple interpretation in terms of so-called Darboux transformations of the orthogonal polynomials \cite{Zhe1}. 
Namely, the formula \eqref{CT_Jac} describes the Christoffel transformation of
the Jacobi polynomials $P_n^{(a,b-1)}(x) \to P_n^{(a,b)}(x)$ while
the formula \eqref{GT_Jac} describes the Geronimus transformation
$P_n^{(a+1,b)}(x) \to P_n^{(a,b)}(x)$.

Combining formulas \eqref{CT_Jac} and \eqref{GT_Jac} we obtain 
\begin{align} 
(1+x)P_n^{(a,b)}(x) &= \alpha_n\,P_{n+1}^{(a+1,b-1)}(x) + \beta_n\,
P_{n}^{(a+1,b-1)}(x) + \gamma_n\, P_{n-1}^{(a+1,b-1)}(x),
\label{1+x_Jac}\\[4pt]
\alpha_n&\eqdef \frac{2(n+1)(n+a+b+1)}{(2n+a+b+1)(2n+a+b+2)},\quad
\beta_n\eqdef \frac{2(a+b)(n+b)}{(2n+a+b)(2n+a+b+2)},\n[2pt]
\gamma_n&\eqdef -\frac{2(n+b)(n+b-1)}{(2n+a+b)(2n+a+b+1)}.\nonumber
\end{align}
Using the formula \eqref{1+x_Jac} with the repeated use of the three term 
recurrence relation \eqref{rec_J} for the factors $\pi(x)$ and $\pi'(x)$
in \eqref{exc_Jac_11},
we conclude that 
\begin{equation}
\hat P_n(x) = \sum_{s=n-\ell}^{n+\ell} \xi_{ns}
P_{s}^{(a+1,b-1)}(x), 
\label{P_P_hat_Jac}
\end{equation}
with some real coefficients
$\xi_{ns}$.

The formula \eqref{P_P_hat_Jac} establishes a ``local" property of the
exceptional polynomials $\hat P_n(x)$ with respect to the basis
$P_{n}^{(a+1,b-1)}(x), \; n=0,1,2,\dots$. Recall that $\deg(\hat P_n(x)) = n + \ell$ and hence degrees of polynomials in lhs and rhs of \eqref{P_P_hat_Jac} coincide.
There is a reciprocal ``local" relation which expresses polynomials
$\pi^2(x) \: P_{n}^{(a+1,b-1)}(x)$ in terms of a finite linear
combination of the exceptional polynomials $\hat P_n(x)$.
In order to get this relation, we can expand the product of
polynomials $\pi^2(x) \: P_{n}^{(a+1,b-1)}(x)$ in terms of the
basis $\hat P_n(x)$: 
\begin{equation}
\pi^2(x) \: P_{n}^{(a+1,b-1)}(x)=
\sum_{s=0}^{\infty} \eta_{ns} \hat P_s(x). 
\label{expan_P_Phat} 
\end{equation}
This sum in general can be infinite, because the polynomials 
$\hat P_n(x)$ form a basis in a Hilbert space with the scalar product given by
the formula \eqref{hat_J_ort} but due to existence of a ``gap" in degrees of
polynomials $\hat P_n(x)$ for a generic polynomial $Q(x)$ we
should have an expansion
$$
Q(x) = \sum_{s=0}^{\infty} \zeta_{ns} \hat P_s(x)
$$
with infinitely many coefficients $\zeta_{ns}$. However, for some
special choices of the polynomial $Q(x)$ this expansion can
contain only a finite number of terms.

In order to find the coefficients $\eta_{ns}$, let us multiply
both sides of \eqref{expan_P_Phat} by $\hat w(x) \hat P_s(x)$ ($\hat w(x)$ is
given by \eqref{w_Jac_b}) and integrate over the interval $[-1,1]$.
Due to the orthogonality relation for polynomials $\hat P_n(x)$, in
the rhs after integration we obtain the term $\hat h_s \eta_{ns}$
with the  nonzero coefficient $\hat h_s$ in  \eqref{hat_J_ort}. On the other hand, in
lhs we have the integral 
\begin{equation}
\int_{-1}^1
(1-x)^{a+1}(1+x)^{b-1} P_{n}^{(a+1,b-1)}(x) \hat P_s(x)dx.
\label{int_P_P_hat} 
\end{equation}
Substituting expression \eqref{P_P_hat_Jac}
into \eqref{int_P_P_hat} and using the orthogonality property of the
Jacobi polynomials $P_{n}^{(a+1,b-1)}(x)$ we have the relation 
\begin{equation}
\hat h_s \eta_{ns} = h_n \xi_{sn}. 
\label{eta_xi_Jac} 
\end{equation}
Hence only
$2 \ell+1$ coefficients $\eta_{ns}$ are nonzero.
We thus have the expansion 
\begin{equation}
 \pi^2(x) \: P_{n}^{(a+1,b-1)}(x)=
\sum_{s=n-\ell}^{n+\ell} \eta_{ns} \hat P_{s}(x),
\label{expan_P_Phat_zeta}
\end{equation}
where the coefficients $\eta_{ns}$ are connected with
$\xi_{ns}$ by the ``mirror" relation \eqref{eta_xi_Jac}.
Formulas \eqref{P_P_hat_Jac} and \eqref{expan_P_Phat_zeta} can be
considered as a generalization for generic $\ell$ of corresponding
formulas obtained for $\ell=1$ in \cite{gomez1}.

Eliminating the Jacobi polynomials $P_{n}^{(a+1,b-1)}(x)$ from
these formulas, we arrive at the recurrence relation 
\begin{equation}
\pi^2(x)
\hat P_n(x) = \sum_{s=n-2 \ell}^{n+2 \ell} K_{ns} \hat P_{n}(x), \quad 
\hat P_{s}(x)=0, \quad \text{if} \ s<0,
\label{rec_exc_Jac} 
\end{equation}
with some real coefficients $K_{ns}$. This
recurrence relation belongs to the class of $4 \ell+1$-diagonal
relations. This means that in the operator form the relation
\eqref{rec_exc_Jac} can be presented as 
\begin{equation}
K {\vec {\hat P}} = \pi^2(x) {\vec {\hat P}}, 
\label{KEVP} 
\end{equation}
where ${\vec {\hat P}}$
is an infinite dimensional vector with components $\hat P_n(x), \;
n=0,1,2,\dots$,
\begin{equation*}
\vec {\hat P} = \{ \hat P_0(x),  \hat P_1(x), \hat P_2(x),
\dots \},
\end{equation*}
and $K$ is a matrix with entries $K_{ns}$. This
matrix has no more than $4 \ell+1$ nonzero diagonals. Corresponding
polynomials $\hat P_n(x)$ satisfy $4 \ell+1$-term recurrence relation
\eqref{rec_exc_Jac}. The ordinary orthogonal polynomials satisfy
3-term recurrence relation. Hence we have polynomials $\hat
P_n(x)$ satisfying more general recurrence relation. Recurrence
relations of such a type were studied e.g. by Dur\'an and Van Assche
\cite{DVA} who showed that such polynomials 
should satisfy a matrix orthogonality relation. Thus the exceptional
Jacobi polynomials belong to the class of polynomials satisfying
higher-order recurrence relations. Note however, that in the
approach of \cite{DVA} it is required that polynomials $\hat
P_n(x)$ have exactly degree $n=0,1,2,\dots$. In our case the
polynomial $\hat P_n(x)$ has degree $n+ \ell, \: n=0,1,2,\dots$. This
means that the methods of \cite{DVA} should be modified if applied to
the case of the exceptional polynomials.

Formulas \eqref{P_P_hat_Jac} and \eqref{expan_P_Phat_zeta} admit an
interesting algebraic interpretation. Introduce semi-infinite
matrices $\Xi$ and $H$ by their entries $\xi_{ns}$ and
$\eta_{ns}$. Then we have 
\begin{equation}
\pi^2(x) \vec P(x) = H \Xi \vec P(x), 
\label{pi_Xi_Eta} 
\end{equation}
where
\begin{equation*}
\vec P(x) = \{P_0^{(a+1,b-1)}(x),  P_1^{(a+1,b-1)}(x),
P_2^{(a+1,b-1)}(x), \dots \},
\end{equation*}
and 
\begin{equation}
\pi^2(x) \vec {\hat P} = \Xi H \vec {\hat P}, \qquad K= \Xi H.
\label{pi_Eta_Xi} 
\end{equation}
{}From the relation \eqref{pi_Xi_Eta} it follows that 
\begin{equation}
\pi^2(J) = H \Xi,
\label{piJ_H_Xi} 
\end{equation}
where $J$ is a Jacobi (3-diagonal) matrix
corresponding to the Jacobi polynomials $P_n^{(a+1,b-1)}(x)$, i.e.
$$
x \vec P(x) = J \vec P(x).
$$
Thus the matrices $H$ and $\Xi$ appear under factorization of the
$4 \ell+1$-diagonal matrix $\pi^2(J)$. The exceptional polynomials $\hat
P_n(x)$ satisfy recurrence relation \eqref{pi_Eta_Xi} which appear
after refactorization (permutation) of the factors $H$ and $\Xi$.
Such permutation of the matrix factors is known as the Darboux
transformation of the matrix $\pi^2(J)$. Similar Darboux
transformations were already studied in \cite{GH2} where
refactorization of the quadratic polynomials in $J$ corresponding
to the Jacobi and Laguerre polynomials was considered. Clearly,
under such refactorization one obtains new polynomials satisfying
a 5-term recurrence relation. These polynomials are not classical
orthogonal polynomials. Nevertheless, the authors of \cite{GH2}
showed that these new polynomials are eigenfunctions of a linear
fourth-order differential operator. In our case we have almost the
same construction as in \cite{GH2} but the resulting polynomials are
exceptional (i.e. some degrees are absent) and  satisfy
a second-order differential equation.

It is also interesting to note that the weight function $\hat w(x)$ given by \eqref{w_Jac_b}, 
differs from the ``classical" weight function $(1-x)^{a+1}(1+x)^{b-1}$ for the Jacobi polynomials 
by the factor $\pi^{-2}(x)$. For the ordinary orthogonal polynomials 
it is known that a rational modification of the weight function $w(x) \to w(x) U(x)^{-1}$ 
with some polynomial $U(x)$ of degree $M$ 
is equivalent to the application of $M$ Geronimus transforms \cite{Zhe1}. 
If $P_n(x)$ are orthogonal polynomials corresponding to the weight function $w(x)$ 
then orthogonal polynomials $\tilde P_n(x)$ corresponding to the weight function 
$w(x) U(x)^{-1}$ are given by the linear combination \cite{Zhe1}
\begin{equation}
\tilde P_n(x)=  A_n^{(0)} P_{n}(x) + A_n^{(1)} P_{n-1}(x) + \dots + A_n^{(M)} P_{n-M}(x),
\label{Ger_PP} 
\end{equation}
with some coefficients $A_n^{(i)}, \: i=0,1,2,\dots M$. 
The formula \eqref{Ger_PP} describes a special case of the $M$-time Geronimus transform 
(general case of the Geronimus transform allows an adding of $M$ concentrated masses 
located on the roots of the polynomial $U(x)$ \cite{Zhe1}).

If we now compare formulas \eqref{P_P_hat_Jac} and \eqref{Ger_PP} 
we see that the exceptional J1 Jacobi polynomials $\hat P_n(x)$ are connected with the ordinary Jacobi polynomials $P_n^{(a+1,b-1)}(x)$ by a transformation which resembles the ordinary $M$-time Geronimus transform, where $M=2 \ell$. 
However, this transformation can be considered as a degeneration of the ordinary Geronimus transform in a sense that the first $\ell$ polynomials $\tilde P_{i}(x), \: i=0,1,\dots, \ell-1$ become zero.

The case of the J2 exceptional Jacobi polynomials can be considered in the same manner leading to a recurrence relation of the same type \eqref{rec_exc_Jac}.

\subsection{Bispectrality of the exceptional Laguerre polynomials }

Consider the case of the exceptional L1 Laguerre polynomials.
Here we list some well known formulas for the ordinary Laguerre polynomials
(see, e.g. \cite{Bateman}): 
\begin{align} 
&\text{Recurrence relation:} \quad -(n+1)L_{n+1}^{(a)}(x)
+ (2n+a+1) L_{n}^{(a)}(x)-(n+a) L_{n-1}^{(a)}(x)\n
&\hspace{60mm}  = x L_{n}^{(a)}(x),
\label{rec_L}  \\
&\text{Differentiation formula:} \qquad\qquad 
{L_{n}^{(a)}}'(x) = - L_{n-1}^{(a+1)}(x),
 \label{diff_L}\\
&\text{Geronimus transformation:}\qquad\ 
L_{n}^{(a)}(x) = L_{n}^{(a+1)}(x) -
L_{n-1}^{(a+1)}(x).
\label{Ger_L}
\end{align}
Following  \eqref{L1xi}, we have $\pi(x) = \xi_\ell(x;g)= L_{\ell}^{(a)}(-x)$, 
where $a\eqdef g+\ell-3/2$. 
We denote the L1 exceptional Laguerre polynomials as $\hat L_n(x)$, 
which are equal to $P_{\ell,n}(x;g)$  given in (O-S2.1)
up to an overall  sign
\begin{equation}
\hat L_n(x) = \pi(x)
{L_{n}^{(a)}}'(x) - (\pi(x) + \pi'(x)) L_{n}^{(a)}(x).
\label{exc_L_j} 
\end{equation}
Here the Geronimus transformation \eqref{Ger_L} is used to rewrite $\xi_\ell(x;g+1)$ in terms of 
$\pi(x)=\xi_\ell(x;g)$.
These polynomials start from degree $\ell$, i.e.
$\deg(\hat L_n(x) ) = n+\ell, \; n=0,1,2,\dots$ and there are no other
members of this orthogonal polynomial family.
The orthogonality relation reads
\begin{equation}
(\hat L_n,\hat L_m)\eqdef \int_{0}^{\infty} \hat L_n(x) \hat L_m
(x) \hat w(x) dx = \hat h_n \delta_{nm}, 
\label{ort_exc_L} 
\end{equation}
where 
\begin{equation}
\hat w(x) = \frac{x^{a+1} e^{-x}}{\pi^2(x)}. 
\label{w_exc_L} 
\end{equation}

Now we
derive recurrence relations for the exceptional Laguerre polynomials.
First of all, using \eqref{diff_L} and \eqref{Ger_L} we rewrite the formula
\eqref{exc_L_j} in the form 
\begin{equation}
\hat L_n(x) =- \pi'(x)L_{n-1}^{(a+1)}(x)  -(\pi'(x) + \pi(x)) L_{n}^{(a+1)}(x).
\label{hat_L_1} 
\end{equation} 
The polynomial $\pi(x)$ has degree $\ell$. Hence, by
the recurrence relation \eqref{rec_L} we have 
\begin{equation}
\hat L_n(x) =
\sum_{s=n-\ell}^{n+\ell} \xi_{ns} L_{s}^{(a+1)}(x), 
\label{hat_L_L} 
\end{equation}
with some real coefficients $\xi_{ns}$. This is a formula expressing
the polynomials $\hat L_n(x)$ as a linear combination of the Laguerre
polynomials $L_{n}^{(a+1)}(x)$.
In order to obtain the reciprocal formula let us consider the
expression
\begin{equation}
\pi^2(x) L_n^{(a+1)}(x) = \sum_{k=0}^{\infty}
\eta_{nk} \hat L_k(x), 
\label{pi_L_eta} 
\end{equation}
 with some real coefficients $\eta_{nk}$. 
Multiply both sides of \eqref{pi_L_eta} by $\hat
L_s(x)$ and integrate with the weight function \eqref{w_exc_L} 
over the interval $(0,\infty)$. 
We then get 
\begin{equation}
\hat h_s \eta_{ns} =
\int_{0}^{\infty} x^{a+1} e^{-x} L_n^{(a+1)}(x) \hat L_s(x) dx.
\label{eta_Lag} 
\end{equation}
Substitute the expression \eqref{hat_L_L} for $\hat
L_s(x)$ in terms of $L_k^{(a+1)}(x)$ into rhs of \eqref{eta_Lag}.
Using the orthogonality relation for the Laguerre polynomials, we have
\begin{equation}
\hat h_s \eta_{ns} = h_n \xi_{sn}. 
\label{Lag_eta_xi} 
\end{equation}
We already have that $\xi_{ns}=0$ if $|n-s|>\ell$. Hence $\eta_{ns}=0$
if $|n-s|>\ell$ and we have the expansion with $2 \ell+1$ terms 
\begin{equation}
\pi^2(x) L_n^{(a+1)}(x) = \sum_{k=n-\ell}^{n+\ell} \eta_{nk} \hat
L_k(x), 
\label{pi_L_eta_f} 
\end{equation} 
where the coefficients $\eta_{ns}$ are connected with $\xi_{ns}$ by \eqref{Lag_eta_xi}.

Similarly to the case of the J1 exceptional Jacobi polynomials we obtain  a $4 \ell+1$-term recurrence relation 
\begin{equation}
\pi^2(x) \hat L_n(x) = \sum_{s=n-2\ell}^{n+2\ell} K_{ns} \hat L_s(x),
\quad 
\hat L_{s}(x)=0, \quad \text{if} \ s<0,
\label{hat_L_rec} 
\end{equation}
The case of the L2 exceptional Laguerre polynomials can be considered in a similar manner leading to a recurrence relation of the same type \eqref{hat_L_rec}.

One can formulate a natural conjecture that all the exceptional orthogonal polynomials 
(i.e. polynomials satisfying a linear second-order Sturm-Liouville equation and 
orthogonal to one another, 
see the detailed description of this problem in \cite{gomez1}) 
satisfy a recurrence relation of the type \eqref{rec_exc_Jac} or \eqref{hat_L_rec} 
with an appropriately defined polynomial $\pi(x)$.

It is worthwhile to stress that the function $\pi(x)^2$ plays the important role of the eigenvalue
of the recurrence relation \eqref{rec_exc_Jac} or \eqref{hat_L_rec}.
In the ordinary three term recurrence relation \eqref{rec_J}  or \eqref{rec_L}, the same role is played
by the (sinusoidal) coordinate $x$ itself.

\subsection{Invariant polynomial subspaces}

Here we will show that the bispectral property of the exceptional Laguerre and Jacobi
polynomials \eqref{rec_exc_Jac}  and \eqref{hat_L_rec} can be easily understood as the characteristic feature of the invariant
polynomial subspaces of the
second order Fuchsian differential operator $\widetilde{\mathcal H}^{\text{O-S}}_\ell$ 
\eqref{htildeeq}.
Because of the denominators containing $\pi(x)\equiv \xi_\ell(x;\bm{\lambda})$, the operator $\widetilde{\mathcal H}^{\text{O-S}}_\ell$ {\em does  not map} a generic polynomial in $x$ into a polynomial.
In other words
\begin{align}
  &\widetilde{\mathcal{H}}^{\text{O-S}}_{\ell}(\bm{\lambda})
  \mathcal{V}_{n}\not\subseteq \mathcal{V}_{n},\qquad
  \mathcal{V}_{n}\eqdef\text{Span}[1,x,\ldots,x^n],  \quad n=0,1,2,\ldots,.
  \label{vnotinv}
\end{align}
However, it is easy to see that
\begin{align}
  &\widetilde{\mathcal{H}}^{\text{O-S}}_{\ell}(\bm{\lambda})
  \pi(x)^2\mathcal{V}_{n}\subseteq \mathcal{V}_{n+2\ell}, \quad
  n=0,1,2,\ldots, .
  \label{vninv}
\end{align}
In other words, $\pi(x)^2 \mathcal{V}_{n}$ belongs to the degree $n+2\ell$ invariant polynomial 
subspace of the operator $\widetilde{\mathcal{H}}^{\text{O-S}}_{\ell}$, see \S6 of \cite{hos}.
Therefore, $\pi(x)^2\hat{P}_n(x)$ and $\pi(x)^2\hat{L}_n(x)$ also  belong to
the degree $n+3\ell$ invariant polynomial subspace and they can be expressed as a linear
combination of $\{\hat{P}_n(x)\}$ ($\{\hat{L}_n(x)\}$):
\begin{equation}
\pi(x)^2\hat{P}_n(x)=\sum_{s=0}^{n+2\ell}K_{ns}\hat{P}_s(x),\quad
\pi(x)^2\hat{L}_n(x)=\sum_{s=0}^{n+2\ell}K_{ns}\hat{L}_s(x),\quad n=0,1,\ldots, .
\end{equation}
Next let us consider the inner products \eqref{hat_J_ort}  and \eqref{ort_exc_L} for $s<n-2\ell$:
\begin{equation*}
(\pi(x)^2\hat{P}_n, \hat{P}_s)=(\hat{P}_n, \pi(x)^2\hat{P}_s),\quad 
(\pi(x)^2\hat{L}_n, \hat{L}_s)=(\hat{L}_n, \pi(x)^2\hat{L}_s).
\end{equation*}
Since
\begin{equation*}
\pi(x)^2\hat{P}_s(x)=\sum_{j=0}^{s+2\ell}K_{sj}\hat{P}_j,\quad
\pi(x)^2\hat{L}_s(x)=\sum_{j=0}^{s+2\ell}K_{sj}\hat{L}_j,
\end{equation*}
we find
\begin{equation}
(\pi(x)^2\hat{P}_n, \hat{P}_s)=0,\quad 
(\pi(x)^2\hat{L}_n, \hat{L}_s)=0,\quad \Longrightarrow 
K_{ns}=0,\quad s<n-2\ell.
\end{equation}
This simply means the desired results
\begin{equation}
\pi(x)^2\hat{P}_n(x)=\sum_{s=n-2\ell}^{n+2\ell}K_{ns}\hat{P}_s(x),\quad
\pi(x)^2\hat{L}_n(x)=\sum_{s=n-2\ell}^{n+2\ell}K_{ns}\hat{L}_s(x),\quad n=0,1,\ldots, .
\end{equation}
This is essentially the same argument for demonstrating the three term recurrence relation.

\section{Comments and discussion}
\label{commentsone}
\setcounter{equation}{0}
A few remarks are in order. Dutta and Roy \cite{duttaroy} derived the first two members of the 
L1 exceptional Laguerre polynomials in a way shown in section \ref{exceplag}.
In our language, they used a prepotential 
\begin{align}
W_\ell(x;g)\eqdef\frac{x^2}{2}+(g+1)\log x+\log u_\ell(x;g),\\
u_\ell(x;g)\eqdef L_\ell^{(g+\frac12)}(-x^2),\quad \ell=1,2.
\end{align}
But erroneously they insist that the partner Hamiltonian 
$\mathcal{H}^{(-)}_\ell(g)$ is {\em not shape invariant\/} on account of the fact that its partner is
the radial oscillator Hamiltonian.

Let us emphasise that the shape invariance is the intrinsic property of the
Hamiltonian, or the potential, determined by the unique factorisation in terms of the groundstate wavefunction. 
Allowing for singularities, a generic quantum mechanical Hamiltonian $\mathcal{H}$ 
has infinitely many different factorisations.
For example, if
\begin{equation}
\mathcal{H}=p^2+V(x),\quad \mathcal{H}\phi_n(x)=\mathcal{E}_n\phi_n(x),
\end{equation}
then it is trivial to show the following factorisation:
\begin{equation}
\mathcal{H}=\left(-\frac{d}{dx}-\frac{\partial_x\phi_n(x)}{\phi_n(x)}\right)
\left(\frac{d}{dx}-\frac{\partial_x\phi_n(x)}{\phi_n(x)}\right)+\mathcal{E}_n,\quad n\ge1,
\end{equation}
The right hand side as a whole is non-singular, but each factor is singular, since $\phi_n(x)$ has $n$ zeros.
Usually the groundstate $n=0$ provides the unique non-singular factorisation.

The results of the present paper assert that the radial oscillator (DPT) potential admits two families,
L1 and L2 (J1 and J2), of {\em infinitely many non-singular factorisations\/}.
Perhaps it would be worthwhile to reflect upon the significance of the infinitely many non-singular factorisations
in historical perspective. More than three decades ago, Miller \cite{miller} embarked on
the program to classify and exhaust factorisations of shape invariant Hamiltonians 
($\mathcal{H}=-d^2/dx^2+V(x)$).
Although he did not employ the word `shape invariance', the essential ingredients were there.
He considers the cases in which the Hamiltonian depends on one parameter only, say, $g$.
Based on an assumption, rephrased in our notation, that the derivative of the prepotential
has a finite power dependence on $g$
\begin{equation}
\frac{dw(x;g)}{dx}=\sum_{j=-N}^M\alpha_j(x) g^j,\quad N, M\in\mathbb{Z}_+,
\end{equation}
he came to the conclusion that the allowed factorisation types were the same as those listed by
Infeld-Hull \cite{infhul}. The present cases of the radial oscillator, for example the L2  \eqref{L2prepot}
\begin{equation}
\text{L2:}\quad \frac{dW_\ell(x;g)}{dx}=-x-\frac{g+\ell}{x}+\frac{\partial_x\xi_\ell(\eta;g)}{\xi_\ell(\eta;g)}
\end{equation}
are not covered by his assumption.

It should be remarked that the present method provides an alternative proof of 
shape invariance of the Odake-Sasaki Hamiltonians \cite{os16, os19, os18, hos}.
Here we will show the shape invariance of the Hamiltonians of the exceptional Laguerre polynomials.
Starting from the prepotential $W_\ell(x;g)$, \eqref{L1prepot} for the L1 and \eqref{L2prepot} for the L2, 
we obtain the Hamiltonian $\mathcal{H}^{(+)}_\ell(g)$, \eqref{radosci2}
for the L1 and \eqref{radosci3} for the L2 and the partner Hamiltonian 
$\mathcal{H}^{(-)}_\ell(g)$, \eqref{partnerradosci2} for the L1 and  \eqref{partnerradosci3} for the L2. 
Then we can construct another partner Hamiltonian of $\mathcal{H}^{(+)}_\ell(g)$ 
by using the factorisation in terms of the groundstate wavefunction
$e^{-x^2/2}x^{g+\ell-1}$ for the L1 and  $e^{-x^2/2}x^{g+\ell+1}$ for the L2.
By the shape invariance of the radial oscillator Hamiltonian, the result
is simply $\mathcal{H}^{(+)}_\ell(g+1)+4$. Thus it is also obtained by the shifted
prepotential $W_\ell(x;g+1)$, which produces the partner Hamiltonian 
$\mathcal{H}^{(-)}_\ell(g+1)$:
\begin{align*}
\begin{CD}
\fbox{\ $\mathcal{H}^{(+)}_\ell(g+1)$\ } 
\quad @>\mbox{\ $W_\ell(x;g+1)$\ }>> \quad
\fbox{\ $\mathcal{H}^{(-)}_\ell(g+1)$\ }\\[4pt]
@A\begin{tabular}{@{}c@{}}\mbox{groundstate}\\ \mbox{wavefunction}\end{tabular}AA
@AA\begin{tabular}{@{}c@{}}\mbox{groundstate}\\ \mbox{wavefunction}\end{tabular}A\\[4pt]
\fbox{\ $\mathcal{H}^{(+)}_\ell(g)$\ } 
\quad @>>\mbox{\ $W_\ell(x;g)$\ }> \quad
\fbox{\ $\mathcal{H}^{(-)}_\ell(g)$\ }\\[4pt]
\begin{tabular}{@{}c@{}}\mbox{Laguerre\quad}\\ \mbox{polynomial\quad}\end{tabular}
@.
\begin{tabular}{@{}c@{}}\mbox{\quad L1, L2 exceptional}\\ 
\mbox{\quad Laguerre polynomial}\end{tabular}
\end{CD}\\[6pt]
\text{Two ways of proving shape invariance}.\hspace{25mm}
\end{align*}
The direct proof of shape invariance of $\mathcal{H}^{(-)}_\ell(g)$ by using the groundstate wavefunction
was given in \cite{os18}.
Obviously the above proof of shape invariance is also valid for the Hamiltonians
of the J1, J2 exceptional Jacobi polynomials, discussed in  section
 \ref{excepjac}.

The annihilation/creation operators of the Hamiltonians of the L1, L2 exceptional Laguerre
polynomials are obtained from those of the corresponding radial oscillator Hamiltonians 
\eqref{radosci2}, \eqref{radosci3},
\begin{equation}
A_\ell(g) a_\ell^{(\pm)} A_\ell^\dagger(g),
\end{equation}
in which $a_\ell^{(\pm)}$ are defined in \eqref{acradosci} with the replacement 
$g\to g+\ell-1$ for the L1 and $g\to g+\ell+1$ for the L2.
Similarly the  annihilation/creation operators of the Hamiltonians of the J1, J2 exceptional Jacobi
polynomials are obtained from those of the corresponding DPT Hamiltonians 
\eqref{PTdown} and \eqref{PTup}
by proper replacements of the parameters.
The creation/annihilation operators for the L1 exceptional Laguerre
polynomials are mentioned in Dutta-Roy paper \cite{duttaroy} for the $\ell=1$ and 2 cases.

The three term recurrence relations for the Laguerre and Jacobi polynomials are
mapped to those of the exceptional orthogonal polynomials simply by $A_\ell$.
The explicit forms are given in \cite{hos}.
This is definitely different from the bispectral property discussed in the previous section 
\ref{sec:bochner}.

\section*{Acknowledgements}

R.\,S. and S.\,T. are supported in part by Grant-in-Aid for Scientific Research from
the Ministry of Education, Culture, Sports, Science and Technology,
 No.22540186 and No.18540214, respectively. A.\,Z. thanks Japan Society for the Promotion of Science for the invitation fellowship.


\end{document}